\pdfoutput=1
\PassOptionsToPackage{unicode}{hyperref}
\PassOptionsToPackage{hyphens}{url}
\PassOptionsToPackage{dvipsnames,svgnames,x11names}{xcolor}
\documentclass[
  10pt,
]{article}
\usepackage{amsmath,amssymb}
\usepackage{setspace}
\usepackage{iftex}
\ifPDFTeX
  \usepackage[T1]{fontenc}
  \usepackage[utf8]{inputenc}
\DeclareUnicodeCharacter{2212}{\ensuremath{-}}
\DeclareUnicodeCharacter{00A0}{~}
  \usepackage{textcomp} % provide euro and other symbols
\else % if luatex or xetex
  \usepackage{unicode-math} % this also loads fontspec
  \defaultfontfeatures{Scale=MatchLowercase}
  \defaultfontfeatures[\rmfamily]{Ligatures=TeX,Scale=1}
\fi

\ifPDFTeX\else
\fi
\IfFileExists{upquote.sty}{\usepackage{upquote}}{}
\IfFileExists{microtype.sty}{% use microtype if available
  \usepackage[]{microtype}
  \UseMicrotypeSet[protrusion]{basicmath} % disable protrusion for tt fonts
}{}
\makeatletter
\@ifundefined{KOMAClassName}{% if non-KOMA class
  \IfFileExists{parskip.sty}{%
    \usepackage{parskip}
  }{% else
    \setlength{\parindent}{0pt}
    \setlength{\parskip}{6pt plus 2pt minus 1pt}}
}{% if KOMA class
  \KOMAoptions{parskip=half}}
\makeatother
\usepackage{xcolor}
\usepackage[a4paper,left=0.95in,right=0.95in,top=0.85in,bottom=0.85in,footskip=0.42in]{geometry}
\usepackage{longtable,booktabs,array}
\usepackage{calc} % for calculating minipage widths
\usepackage{etoolbox}
\makeatletter
\patchcmd\longtable{\par}{\if@noskipsec\mbox{}\fi\par}{}{}
\makeatother
\IfFileExists{footnotehyper.sty}{\usepackage{footnotehyper}}{\usepackage{footnote}}
\makesavenoteenv{longtable}
\usepackage{graphicx}
\makeatletter
\def\maxwidth{\ifdim\Gin@nat@width>\linewidth\linewidth\else\Gin@nat@width\fi}
\def\maxheight{\ifdim\Gin@nat@height>\textheight\textheight\else\Gin@nat@height\fi}
\makeatother
\setkeys{Gin}{width=\maxwidth,height=\maxheight,keepaspectratio}
\makeatletter
\def\fps@figure{htbp}
\makeatother
\usepackage{xurl}
\usepackage{xcolor}
\definecolor{srblue}{HTML}{0000EE}
\definecolor{srrule}{HTML}{000000}
\usepackage{mathptmx}
\DeclareMathAlphabet{\mathcal}{OMS}{cmsy}{m}{n}
\DeclareFontFamily{T1}{phvsr}{}
\DeclareFontShape{T1}{phvsr}{m}{n}{<-> s*[0.859] phvr8t}{}
\DeclareFontShape{T1}{phvsr}{m}{sl}{<-> s*[0.859] phvro8t}{}
\DeclareFontShape{T1}{phvsr}{m}{it}{<->ssub * phvsr/m/sl}{}
\DeclareFontShape{T1}{phvsr}{b}{n}{<-> s*[0.835] phvb8t}{}
\DeclareFontShape{T1}{phvsr}{b}{sl}{<-> s*[0.835] phvbo8t}{}
\DeclareFontShape{T1}{phvsr}{b}{it}{<->ssub * phvsr/b/sl}{}
\DeclareFontShape{T1}{phvsr}{bx}{n}{<->ssub * phvsr/b/n}{}
\DeclareFontShape{T1}{phvsr}{bx}{it}{<->ssub * phvsr/b/sl}{}
\DeclareFontFamily{TS1}{phvsr}{}
\DeclareFontShape{TS1}{phvsr}{m}{n}{<-> s*[0.859] phvr8c}{}
\DeclareFontShape{TS1}{phvsr}{m}{sl}{<-> s*[0.859] phvro8c}{}
\DeclareFontShape{TS1}{phvsr}{m}{it}{<->ssub * phvsr/m/sl}{}
\DeclareFontShape{TS1}{phvsr}{b}{n}{<-> s*[0.835] phvb8c}{}
\DeclareFontShape{TS1}{phvsr}{b}{sl}{<-> s*[0.835] phvbo8c}{}
\DeclareFontShape{TS1}{phvsr}{b}{it}{<->ssub * phvsr/b/sl}{}
\DeclareFontShape{TS1}{phvsr}{bx}{n}{<->ssub * phvsr/b/n}{}
\DeclareFontShape{TS1}{phvsr}{bx}{it}{<->ssub * phvsr/b/sl}{}
\newcommand{\srsans}{\fontfamily{phvsr}\selectfont}
\usepackage{lastpage}
\usepackage{booktabs}
\usepackage{microtype}
\usepackage{needspace}
\usepackage[labelformat=empty,font={small},justification=justified,singlelinecheck=false,skip=6pt]{caption}
\usepackage{fancyhdr}
\fancypagestyle{plain}{\fancyhf{}%
  \fancyfoot[R]{\srsans\small\bfseries\thepage/\pageref*{LastPage}}}

\newcommand{\srmedium}{\fontfamily{phvsr}\selectfont\lsstyle}
\definecolor{oiblue}{HTML}{0F52BA}
\newenvironment{tkfeyebrow}{\par\setlength{\parindent}{0pt}\srmedium\bfseries\fontsize{9}{11}\selectfont\color{oiblue}}{\par\vspace{6pt}}
\newenvironment{tkftitle}{\par\setlength{\parindent}{0pt}\srsans\bfseries\fontsize{22}{26}\selectfont\raggedright}{\par\vspace{16pt}}

\newenvironment{tkfauthor}{\par\setlength{\parindent}{0pt}\srsans\bfseries\fontsize{13}{16}\selectfont}{\par\vspace{1pt}}
\newenvironment{tkfaffil}{\par\setlength{\parindent}{0pt}\srsans\fontsize{10}{13}\selectfont\color{srgrey}\hypersetup{urlcolor=srgrey}}{\par\vspace{12pt}}
\definecolor{srgrey}{HTML}{6B6B6B}

\newlength{\srgap}
\newenvironment{tkfabshead}{\par\setlength{\parindent}{0pt}\srsans\bfseries\fontsize{14}{17}\selectfont}{\par\vspace{4pt}}
\newenvironment{tkfgraphical}{\par\vspace{39pt}\setlength{\parindent}{0pt}\begin{list}{}{\leftmargin=0pt\rightmargin=0pt\topsep=0pt}\item[]}{\end{list}\par\vspace{\srgap}}
\newenvironment{tkfabstract}{\par\setlength{\parindent}{0pt}\begin{list}{}{\leftmargin=0pt\rightmargin=0pt\topsep=0pt}\item[]\srsans\fontsize{10}{13}\selectfont}{\end{list}\par\vspace{\srgap}}
\newenvironment{tkfkeywords}{\par\setlength{\parindent}{0pt}\begin{list}{}{\leftmargin=0pt\rightmargin=0pt\topsep=0pt}\item[]\srsans\fontsize{10}{13}\selectfont\raggedright}{\end{list}\par\vspace{\srgap}}
\newenvironment{tkfsec}{\par\Needspace{4\baselineskip}\vspace{12pt}\noindent\srsans\bfseries\fontsize{14}{17}\selectfont\raggedright\nopagebreak}{\par\nopagebreak\vspace{4pt}\noindent\ignorespaces}
\newenvironment{tkfsubsec}{\par\Needspace{4\baselineskip}\vspace{8pt}\noindent\srsans\bfseries\fontsize{11}{13.5}\selectfont\raggedright\nopagebreak}{\par\nopagebreak\vspace{1pt}}
\newenvironment{tkfsubsecfirst}{\par\nopagebreak\noindent\srsans\bfseries\fontsize{11}{13.5}\selectfont\raggedright\nopagebreak}{\par\nopagebreak\vspace{1pt}}
\newenvironment{tkfref}{\par\small\linespread{1.05}\selectfont\raggedright\setlength{\parindent}{0pt}\hypersetup{urlcolor=black}\everypar{\hangindent=0.5in\hangafter=1}}{\par\vspace{2pt}}
\newenvironment{tkftablenote}{\par\noindent\small\linespread{1.0}\selectfont}{\par\vspace{12pt}}
\newenvironment{tkftabletitle}{\par\Needspace{8\baselineskip}\vspace{6pt}\setlength{\parindent}{0pt}\small}{\par\vspace{-2pt}}
\newenvironment{tkfappendix}{\par\setlength{\parindent}{0pt}\srsans\bfseries\fontsize{14}{17}\selectfont}{\par\vspace{6pt}}
\AtBeginEnvironment{longtable}{\small\linespread{1.05}\selectfont}


\AtBeginDocument{\raggedbottom}
\usepackage{etoolbox}
\makeatletter\AfterEndEnvironment{tkfsec}{\@afterindentfalse\@afterheading}\AfterEndEnvironment{tkfsubsec}{\@afterindentfalse\@afterheading}\AfterEndEnvironment{tkfsubsecfirst}{\@afterindentfalse\@afterheading}\makeatother

\newcommand{\srcorresponding}[1]{\begingroup\renewcommand\thefootnote{}\footnotetext{\hspace{-1.8em}{\srsans #1}}\endgroup}
\ifLuaTeX
  \usepackage{selnolig}  % disable illegal ligatures
\fi
\IfFileExists{bookmark.sty}{\usepackage{bookmark}}{\usepackage{hyperref}}
\IfFileExists{xurl.sty}{\usepackage{xurl}}{} % add URL line breaks if available
\hypersetup{
  colorlinks=true,
  linkcolor={srblue},
  filecolor={Maroon},
  citecolor={Blue},
  urlcolor={Blue},
  pdfcreator={LaTeX via pandoc}}

\author{}
\date{}

\begin{document}

\setstretch{1.05}
\begin{tkfeyebrow}

RESEARCH LETTER

\end{tkfeyebrow}

\begin{tkftitle}

Quantum amplification of improbable futures

\end{tkftitle}

\begin{tkfauthor}

Peter Maurice Catt

\end{tkfauthor}

\begin{tkfaffil}

Auckland, New Zealand ·
\href{https://www.theknowablefuture.com}{www.theknowablefuture.com} ·
ORCID 0009-0000-1324-0861

\end{tkfaffil}

\srcorresponding{Corresponding author: peter.catt@theknowablefuture.com}

\begin{tkfabshead}

ABSTRACT

\end{tkfabshead}

\begin{tkfabstract}

Some consequential futures consist of entirely plausible steps arranged
into an unlikely sequence, such as ordinary weather anomalies combining
into an extreme event or local equipment failures propagating into a
widespread outage. Because the sequence as a whole carries little
probability, it may be rarely observed in simulation. Quantum mechanics
offers a way to change those odds: ideal amplitude amplification raises
the chance of drawing such futures while preserving their relative
likelihoods. For a five-state process six steps ahead, 13,888 of 15,625
futures are rare, yet together they account for about one draw in 40. We
built the complete circuit in IBM's Qiskit and checked every part
against an exact enumeration; four simulated rounds raised the share of
rare futures drawn from 2.5\% to 98.0\%. On IBM's Kingston quantum
computer, a two-step reference circuit that marks rare futures from a
precomputed list drew them in 68.4\% of shots against 43.5\% for its
control (\(z=50\)); at three steps (about 1,170 two-qubit gates) the
predicted signal reversed sign. Because this model's probabilities are
known, every result could be checked against exact values; the mechanism
is intended for simulators whose probabilities are hard to calculate,
such as climate, epidemic and financial models.

\end{tkfabstract}

\begin{tkfkeywords}


\end{tkfkeywords}

\begin{tkfgraphical}\includegraphics[width=\linewidth]{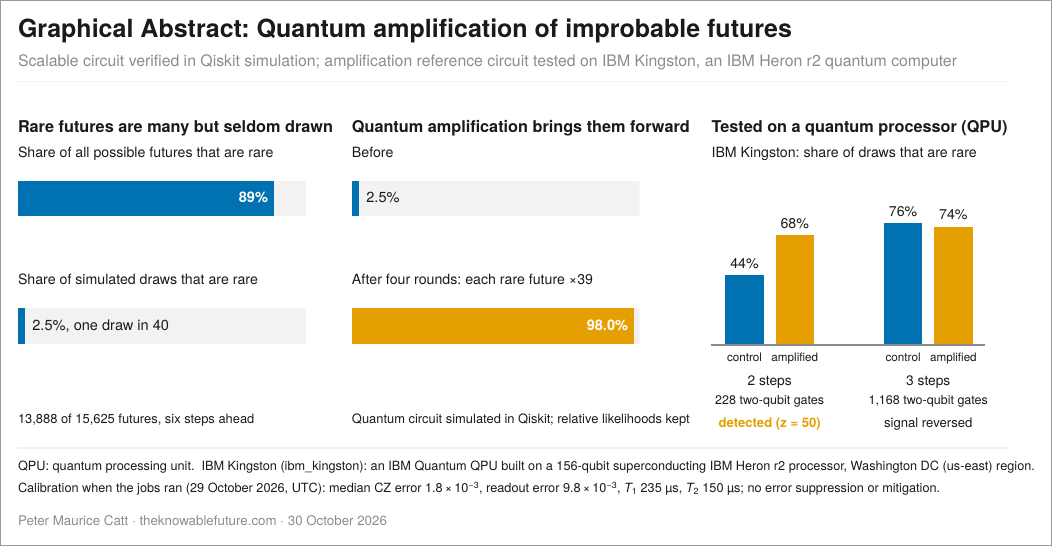}\end{tkfgraphical}

\clearpage

\begin{tkfsec}

Introduction

\end{tkfsec}

Many consequential outcomes are path dependent. Droughts and floods can
arise from combinations of drivers that are individually unremarkable
(Zscheischler et~al., 2018), financial distress spreads through chains
of exposure between institutions (Haldane \& May, 2011), and power-grid
blackouts propagate through sequences of individual line failures
(Dobson et~al., 2007). Classical rare-event simulation, through
importance sampling, multilevel splitting and genealogical particle
methods, is highly effective once the event of interest has been
specified (Asmussen \& Glynn, 2007; Bucklew, 2004; Cérou \& Guyader,
2007; Del Moral \& Garnier, 2005; Dupuis \& Wang, 2005). Guo
et~al.~(2026) consider a different target: every outcome whose own
probability is at most a threshold \(\Delta\), without naming those
outcomes in advance. For stationary processes they set
\(\Delta=2^{-\alpha L}\) for trajectories of length \(L\), with
\(\alpha\) at least the entropy rate, so that a trajectory is rare when
its surprisal per step is at least \(\alpha\). They show that a
classical algorithm with sample access needs \(\Omega(1/\Delta)\)
samples to sample from the rare-conditioned distribution, whereas a
quantum algorithm with coherent access to a state-preparation unitary
and its inverse needs
\(\widetilde{\mathcal O}\big(1/\sqrt{p_{\mathrm{rare}}\Delta}\big)\)
applications of them, where \(p_{\mathrm{rare}}\) is the total
probability of the rare set.

Guo et~al.~(2026) build the marking of rare outcomes from the
state-preparation unitary itself, by applying a polynomial threshold to
outcome probabilities, so that rare outcomes never have to be
identified. When the process is known, the marking can instead be
computed from its rules, and to run amplitude amplification (Brassard
et~al., 2002; Grover, 1996) on a quantum computer that computation has
to be a circuit. Amplitude amplification and estimation already underpin
quantum approaches to Monte Carlo estimation and financial risk (Egger
et~al., 2021; Montanaro, 2015; Stamatopoulos et~al., 2020; Woerner \&
Egger, 2019), and Grover search has been run on hardware at a few qubits
(Figgatt et~al., 2017). For a Markov process with a known transition
matrix, the surprisal of a trajectory is a sum of per-step terms, so it
can be computed coherently and compared with the threshold. This letter
builds that circuit at gate level in Qiskit, verifies it against exact
enumeration, records its logical cost as the horizon grows, and tests
the amplification step on IBM superconducting hardware.

Because the circuit is built from the transition matrix, a classical
computer holding the same matrix can compute the rare-set probability,
and sample from the rare-conditioned distribution, exactly by dynamic
programming over states and accumulated fixed-point surprisal. The
contribution is a verified, costed gate-level implementation of the
amplification of improbable futures, with a hardware test of its central
step.

\begin{tkfsec}

Methods

\end{tkfsec}

\begin{tkfsubsecfirst}

Stochastic process and future trajectories

\end{tkfsubsecfirst}

The process has five states and transition matrix

{\small\setlength{\arraycolsep}{4pt}\begin{equation}T=\begin{pmatrix}0.88&0.05&0.03&0.025&0.015\\0.20&0.65&0.07&0.05&0.03\\0.15&0.08&0.62&0.10&0.05\\0.10&0.08&0.12&0.62&0.08\\0.08&0.06&0.10&0.12&0.64\end{pmatrix}.\tag{1}\end{equation}}\vspace{8pt}

Its stationary entropy rate is \(h_\infty=1.1535596\) bits per step: the
uncertainty in each step that remains once the full past is known. For a
stationary process it equals the entropy of a single step minus the
one-step forecastability, the mutual information between the next step
and the full past (Catt, 2026; Cover \& Thomas, 2006). A future
\(x=(x_1,\ldots,x_H)\) is the sequence of the next \(H\) states given
\(X_0=1\), with probability \(P_H(x)=\prod_{t=1}^{H}T_{x_{t-1}x_t}\).
Following Guo et~al.~(2026) with \(\alpha=h_\infty+\delta\), the rare
set, its probability and the threshold are

\begin{equation}\begin{gathered}\mathcal R_\Delta=\left\{x:0<P_H(x)\le\Delta_H\right\},\\ p_{\mathrm{rare}}=\sum_{x\in\mathcal R_\Delta}P_H(x),\qquad \Delta_H=2^{-H(h_\infty+\delta)},\end{gathered}\tag{2}\end{equation}

so a future is rare when its surprisal \(-\log_2P_H(x)\) is at least
\(S_\Delta=H(h_\infty+\delta)\); no future lies exactly on the
threshold, so the choice between a strict and a non-strict inequality
does not matter here. We fix \(\delta=1.5\) bits per step throughout.
Guo et~al.~treat stationary processes; here every future instead starts
in state 1, the most persistent state, so the threshold is a fixed
reference level set by the stationary entropy rate, not an estimate of
the expected surprisal from state 1.

\begin{tkfsubsec}

Encoding and state preparation

\end{tkfsubsec}

Each step is held in three qubits, of which five basis states encode the
process states and three are unused, so the path register has \(3H\)
qubits. The trajectory distribution is prepared as

\begin{equation}\vert P_H\rangle=\sum_x\sqrt{P_H(x)}\,\vert x\rangle\tag{3}\end{equation}

by a cascade of uniformly controlled \(R_y\) rotations (Grover \&
Rudolph, 2002; Möttönen et~al., 2004, 2005). Within each step the three
bits are prepared from most to least significant, each conditioned on
the previous step's register and on the bits already prepared. Rotations
conditioned on unused encodings are set to zero, so the prepared state
has no amplitude on invalid encodings. At \(H=6\) the prepared
probabilities matched the enumerated \(P_6(x)\) to within
\(2.4\times10^{-15}\) in every basis state.

\begin{tkfsubsec}

Fixed-point surprisal

\end{tkfsubsec}

Transition surprisals and the threshold are held to \(b\) fractional
bits,

\begin{equation}s_{ij}=\operatorname{round}\left(2^{b}\left(-\log_2T_{ij}\right)\right),\qquad T_b=\left\lceil2^{b}S_\Delta\right\rceil.\tag{4}\end{equation}

The fixed-point test \(\sum s\ge T_b\) can disagree with the real-valued
definition for futures close to the threshold. Rounding moves each of
the \(H\) transition surprisals by at most \(2^{-(b+1)}\) bits and the
threshold by at most \(2^{-b}\), so no future can be misclassified once
\((H/2+1)\,2^{-b}\) is less than the gap between the threshold and the
nearest future; below that bound, agreement depends on how the rounding
errors fall and has to be checked. Two kinds of error are therefore
recorded separately: a circuit error, where the circuit's result differs
from the fixed-point test, which must be zero; and a precision error,
where the fixed-point test differs from the real-valued rare set,
reported as a count and as misclassified probability mass.

\begin{tkfsubsec}

Rarity oracle

\end{tkfsubsec}

An accumulator register of \(m\) qubits is initialised to an offset
\(c\), transformed to the Fourier basis, and receives each transition
surprisal as phase rotations (Draper, 2000; Ruiz-Perez \&
Garcia-Escartin, 2017): for accumulator qubit \(j\), a diagonal gate
(Shende et~al., 2006) on that qubit and the step registers the
transition depends on applies the phase \(2\pi s/2^{j+1}\). The inverse
transform then leaves \(c+\sum s \bmod 2^m\) in the register. With

\begin{equation}\begin{gathered}2^{m-1}\ge\max\left(T_b,\;S_{\max}-T_b+1\right),\\ c=2^{m-1}-T_b,\end{gathered}\tag{5}\end{equation}

where \(S_{\max}\) is the largest fixed-point surprisal over valid
futures, every sum stays in range and the most significant bit equals 1
exactly when \(\sum s\ge T_b\). The oracle applies \(Z\) to that bit and
uncomputes the accumulator, so no comparator and no validity flag are
needed. Unused encodings are given zero surprisal; they never carry
amplitude, because the preparation assigns them none and both the oracle
and the reflection about \(|P_H\rangle\) preserve the span of valid
basis states.

\begin{tkfsubsec}

Amplitude amplification

\end{tkfsubsec}

With \(\sin^2\theta=p\), the probability of the marked set, \(r\)
iterations of the amplitude-amplification operator give (Brassard
et~al., 2002; Grover, 1996)

\begin{equation}P_r=\sin^2\left[(2r+1)\theta\right],\qquad r^{*}=\left\lfloor\frac{\pi}{4\theta}\right\rfloor.\tag{6}\end{equation}

Each iteration applies the oracle, the inverse preparation, a reflection
about the all-zero state of the path register and the preparation. Every
marked amplitude is scaled by the same factor, so the distribution
within the marked set is preserved. The iteration count \(r^*\) is that
of Brassard et~al.~(2002); at every horizon studied here it is also the
first integer \(r\) at which \(P_r\) reaches a local maximum.

\begin{tkfsubsec}

Validation

\end{tkfsubsec}

Three checks were applied at \(H=6\). First, for every one of the 15,625
valid futures, the path register was set to that basis state, the
accumulator circuit was run in Qiskit Aer's matrix-product-state
simulator (Vidal, 2003), and the full accumulator register was read and
compared with \(c+\sum s\bmod 2^m\). Second, the complete amplification
circuit, from preparation through four iterations, was simulated as a
statevector at \(b=7\) and at \(b=4\), and after each iteration the
probability of every path-register basis state and of the accumulator's
zero state was recorded and compared with exact amplitude amplification
of the enumerated distribution under the same fixed-point marking.
Third, the probability of a rare future under the real-valued definition
was computed directly from the recorded probabilities, never inferred
from the analytical formula.

\begin{tkfsubsec}

Resource accounting

\end{tkfsubsec}

Logical resources were counted after transpiling the same circuits to
the basis \(\{R_z,\sqrt X,X,\mathrm{CNOT}\}\) with all-to-all
connectivity at optimisation level 3. Counts are reported at a fixed
precision, \(b=4\), and at the lowest \(b\) with zero mismatches.

\begin{tkfsubsec}

Hardware reference and matched control

\end{tkfsubsec}

The full circuit is too deep for present hardware, so the amplification
step was tested with compact references at \(H=2\) and \(H=3\). Each
indexes complete futures in base 5, using 5 and 7 qubits respectively,
prepares the trajectory distribution with a general state-preparation
circuit, and marks rare futures with a diagonal gate built from the
enumerated rare set. These references serve only to test amplification
on hardware and do not scale. Because rare futures are the majority by
count (21 of 25 at \(H=2\), 109 of 125 at \(H=3\)), noise that flattens
the output distribution raises the raw rare-hit rate on its own. The
primary statistic is therefore the difference between each circuit and a
matched control: the same circuit with a diagonal gate that is the
identity on all valid futures and has random phases on unused indices,
compiled to the same physical qubits. An identity would be removed by
the compiler; the random phases, on indices that carry no amplitude in
the ideal circuit, make it synthesise a diagonal gate on the same
qubits, of comparable routed two-qubit cost. Noisy simulations used
noise models built from IBM calibration snapshots of Fez, Kingston and
Marrakesh, retrieved on 29 October 2026, with 3 repeats of 20,000 shots
per arm.

The hardware test was fixed before any run: a one-sided, pooled
two-proportion \(z\) test of oracle against control, passed when
\(z\ge5\), using four alternating oracle and control pairs of 5,000
shots each, giving 20,000 shots per arm, submitted as eight primitive
unified blocs (PUBs) of a single Sampler job for each horizon, on the
backend chosen from the noisy-simulation results before submission.
Control and oracle circuits must agree in routed two-qubit count to
within 2\%.

\begin{tkfsubsec}

Software and hardware

\end{tkfsubsec}

Circuits were built, simulated and compiled for IBM hardware with Qiskit
(Javadi-Abhari et~al., 2024) and Qiskit Aer, and submitted with Qiskit
IBM Runtime version 0.51.0. The 30-qubit statevector simulation at
\(b=7\) used Qiskit version 2.5.2 and Qiskit Aer version 0.17.2 on an
Apple-silicon Mac with 32 GB of memory and took 39 minutes; the
calibration-snapshot noisy simulations, and the compilation and
submission of the hardware jobs, used the same software on the same
machine. All other computations used Qiskit version 2.6.1 and Qiskit Aer
version 0.18.2 on two CPU cores of a Linux machine, where the 27-qubit
simulation at \(b=4\) took 17 minutes. NumPy computed the classical
reference values: trajectory probabilities, the rare set and its mass.
Hardware runs used IBM Kingston (ibm\_kingston), a quantum processing
unit (QPU) on the IBM Quantum Platform built on a 156-qubit
superconducting IBM Heron r2 processor and hosted in the platform's
Washington DC (us-east) region. IBM Fez and Marrakesh, QPUs with the
same processor type, were used only in the noisy simulations that
selected it. Hardware circuits were executed with the Sampler
primitive's default options, under which no dynamical decoupling,
twirling or readout mitigation is applied, so the measured signal
reflects the processor's native noise. In the calibration in force when
the jobs ran, Kingston's median CZ error was \(1.8\times10^{-3}\), its
median readout error \(9.8\times10^{-3}\), and its median \(T_1\) and
\(T_2\) were 235 and 150 µs; values for the qubits used are given with
Table 2.

\begin{tkfsec}

Results

\end{tkfsec}

\begin{tkfsubsecfirst}

Rare sets and precision

\end{tkfsubsecfirst}

At \(H=6\), 13,888 of the 15,625 futures are rare, yet together they
carry probability 0.0250 (Supplementary Table S1). The nearest future
lies \(5.7\times10^{-4}\) bits from the threshold, so the bound above
guarantees exact classification only from \(b=13\). In practice \(b=7\)
already gives zero mismatches, while \(b=4\) misclassifies 9 futures
carrying \(1.4\times10^{-4}\) of probability, 0.57\% of the rare mass
(Supplementary Table S2). Below the guaranteed bound agreement is not
monotone in \(b\): \(b=9\) misclassifies two futures even though \(b=7\)
and \(b=8\) misclassify none, so exactness at \(b=7\) is a verified
property of that choice, not a guaranteed one.

\begin{tkfsubsec}

Circuit validation

\end{tkfsubsec}

The accumulator returned the exact register value on all 15,625 valid
futures at \(b=7\) and on all 15,625 at \(b=4\), with no errors. With
Supplementary Table S2 and the statevector check below, this means the
\(b=7\) oracle marks exactly the real-valued rare set and the \(b=4\)
oracle marks exactly its fixed-point set.

\begin{tkfsubsec}

Amplification at six steps

\end{tkfsubsec}

Statevector simulation of the complete 30-qubit circuit at \(b=7\)
followed exact amplitude amplification of the real-valued rare set to
within \(5.9\times10^{-15}\) in every trajectory probability at every
iteration (Fig. 1, Table 1). The accumulator returned to zero to within
\(4.2\times10^{-14}\), and the probability on unused encodings stayed
below \(2.5\times10^{-31}\). After four iterations a rare future was
drawn with probability 0.9798, against 0.0250 before amplification, and
the amplified rare-conditioned distribution lay within
\(5.0\times10^{-15}\) of its target in total variation, so the relative
likelihoods of the rare futures were unchanged.

The 27-qubit simulation at \(b=4\) followed exact amplification of its
fixed-point marked set equally closely, to within \(4.4\times10^{-15}\).
Measured against the real-valued target, its final probability was
0.9787 and its rare-conditioned distribution lay \(5.7\times10^{-3}\)
from the target, the effect of the 9 futures misclassified at \(b=4\),
not a circuit error.

\begin{figure}
\centering
\includegraphics[width=10cm,height=\textheight]{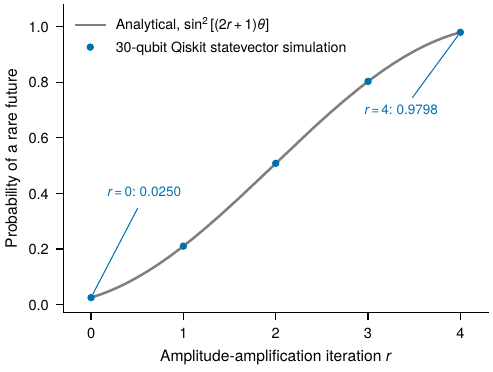}
\caption{{\srsans\bfseries {Figure 1.}} Probability of drawing a rare
future against amplification iteration \(r\) at \(H=6\) (\(\delta=1.5\),
\(b=7\)). The grey line is \(\sin^2[(2r+1)\theta]\) with
\(\sin^2\theta=p_{\mathrm{rare}}\). Blue points are the 30-qubit Qiskit
statevector simulation, computed directly from the simulated
probabilities of all \(2^{18}\) path-register states; they lie on the
curve to within \(4.1\times10^{-14}\).}
\end{figure}

\begin{tkfsubsec}

Logical resources

\end{tkfsubsec}

At fixed \(b=4\) the oracle grows from 2,441 to 11,777 two-qubit gates
between \(H=2\) and \(H=6\) (Supplementary Table S3). The cost is
dominated by the diagonal gates that couple consecutive step registers
to an accumulator qubit, \((H-1)m\) in the computation and as many again
in the uncomputation, each on seven qubits and compiled to \(2^7-2=126\)
CNOTs (Shende et~al., 2006). With the Fourier transforms and the first
step, the oracle costs 262 to 305 two-qubit gates per transition and
accumulator qubit, so it grows in proportion to \((H-1)m\); the step
between \(H=3\) and \(H=4\) reflects the accumulator growing from 8 to 9
qubits. Exact marking changes the picture through the accumulator width:
at \(H=5\) the lowest \(b\) with zero mismatches is 13, giving an oracle
of 19,725 two-qubit gates, more than at \(H=6\) with \(b=7\) (15,933).

\begin{tkfsubsec}

Hardware reference

\end{tkfsubsec}

Noise erodes much of the ideal signal at two steps and most of it at
three (Fig. 2, Supplementary Table S4). At \(H=2\) with one iteration
the predicted signal is +0.35 to +0.41 against an ideal +0.65, with
\(z\) from 70 to 82. At \(H=3\) it falls to +0.039 to +0.060 against an
ideal +0.52, with \(z\) from 8.6 to 12.5. On Fez and Marrakesh the
\(H=3\) control missed the 2\% matching tolerance, so only Kingston
gives a valid three-step comparison, with a predicted \(z\) of 12.5. A
second iteration raises the ideal success but lowers the predicted
signal at \(H=2\) on all three backends, and at \(H=3\) on all three,
although every three-step control at \(r=2\) misses the matching rule.
One iteration is therefore the preferred design at both horizons, and
Kingston, which gives the largest predicted signal at both, is the
chosen backend. The raw rare-hit rates show why the control is required:
the control alone reaches well above \(p_{\mathrm{rare}}\) because noise
moves probability towards the many rare futures, which make up 0.656 of
a uniform output over all indices at \(H=2\) and 0.852 at \(H=3\).

\Needspace{24\baselineskip}

\begin{tkftabletitle}

{\srsans\bfseries {Table 1.}} Amplification at \(H=6\) (Qiskit Aer
statevector)

\end{tkftabletitle}

\begin{longtable}[]{@{}
  >{\raggedleft\arraybackslash}p{(\columnwidth - 12\tabcolsep) * \real{0.0930}}
  >{\raggedleft\arraybackslash}p{(\columnwidth - 12\tabcolsep) * \real{0.0930}}
  >{\raggedleft\arraybackslash}p{(\columnwidth - 12\tabcolsep) * \real{0.1628}}
  >{\raggedleft\arraybackslash}p{(\columnwidth - 12\tabcolsep) * \real{0.1628}}
  >{\raggedleft\arraybackslash}p{(\columnwidth - 12\tabcolsep) * \real{0.1628}}
  >{\raggedleft\arraybackslash}p{(\columnwidth - 12\tabcolsep) * \real{0.1628}}
  >{\raggedleft\arraybackslash}p{(\columnwidth - 12\tabcolsep) * \real{0.1628}}@{}}
\toprule\noalign{}
\begin{minipage}[b]{\linewidth}\raggedleft
\(b\)
\end{minipage} & \begin{minipage}[b]{\linewidth}\raggedleft
\(r\)
\end{minipage} & \begin{minipage}[b]{\linewidth}\raggedleft
\(P(\mathrm{rare})\), real definition
\end{minipage} & \begin{minipage}[b]{\linewidth}\raggedleft
Exact amplification, fixed-point marking
\end{minipage} & \begin{minipage}[b]{\linewidth}\raggedleft
Max error per trajectory
\end{minipage} & \begin{minipage}[b]{\linewidth}\raggedleft
Accumulator leakage
\end{minipage} & \begin{minipage}[b]{\linewidth}\raggedleft
\(D_{\mathrm{TV}}\) to real-valued target
\end{minipage} \\
\midrule\noalign{}
\endhead
\bottomrule\noalign{}
\endlastfoot
7 & 0 & 0.024977 & 0.024977 & \(2.3\times10^{-15}\) & n/a &
\(4.6\times10^{-15}\) \\
7 & 1 & 0.210066 & 0.210066 & \(5.9\times10^{-15}\) &
\(1.3\times10^{-14}\) & \(4.6\times10^{-15}\) \\
7 & 2 & 0.508127 & 0.508127 & \(3.8\times10^{-15}\) &
\(1.9\times10^{-14}\) & \(4.7\times10^{-15}\) \\
7 & 3 & 0.803021 & 0.803021 & \(1.4\times10^{-15}\) &
\(3.2\times10^{-14}\) & \(4.7\times10^{-15}\) \\
7 & 4 & 0.979845 & 0.979845 & \(5.3\times10^{-16}\) &
\(4.1\times10^{-14}\) & \(4.9\times10^{-15}\) \\
4 & 0 & 0.024977 & 0.024977 & \(1.1\times10^{-15}\) & n/a &
\(4.6\times10^{-15}\) \\
4 & 1 & 0.209060 & 0.209060 & \(4.3\times10^{-15}\) &
\(6.0\times10^{-15}\) & \(5.2\times10^{-3}\) \\
4 & 2 & 0.505902 & 0.505902 & \(7.5\times10^{-16}\) &
\(1.1\times10^{-15}\) & \(5.6\times10^{-3}\) \\
4 & 3 & 0.800485 & 0.800485 & \(1.9\times10^{-17}\) &
\(1.5\times10^{-14}\) & \(5.7\times10^{-3}\) \\
4 & 4 & 0.978669 & 0.978669 & \(7.2\times10^{-16}\) &
\(2.7\times10^{-14}\) & \(5.7\times10^{-3}\) \\
\end{longtable}

\begin{tkftablenote}

\(b=7\) uses 30 qubits and marks the real-valued rare set exactly;
\(b=4\) uses 27 qubits and misclassifies 9 futures. \(P(\mathrm{rare})\)
is computed directly from the simulated probabilities of all \(2^{18}\)
path-register states. The exact-amplification column gives
\(P(\mathrm{rare})\), real definition, under exact amplification of the
fixed-point marked set. Accumulator leakage is the magnitude of
\(1-P(\text{accumulator}=0)\). \(D_{\mathrm{TV}}\) is the
total-variation distance of the amplified rare-conditioned distribution
from its real-valued target.

\end{tkftablenote}

\begin{figure}
\centering
\includegraphics[width=10cm,height=\textheight]{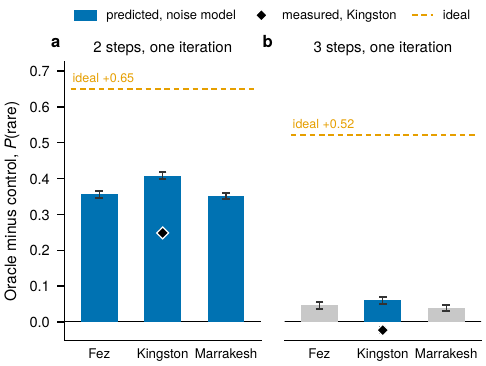}
\caption{{\srsans\bfseries {Figure 2.}} Oracle-minus-control signal for
one amplification iteration, predicted under noise models built from IBM
calibration snapshots retrieved on 29 October 2026, and measured on
Kingston. {\srsans\bfseries {a}}, \(H=2\) (5 qubits).
{\srsans\bfseries {b}}, \(H=3\) (7 qubits). Bars show means over 3
simulations of 20,000 shots per arm; error bars show 95\% binomial
intervals for a 20,000-shot hardware run. Grey bars fall below the
\(z\ge5\) threshold or miss the 2\% matching rule. The dashed line is
the ideal signal, \(\sin^2 3\theta-p_{\mathrm{rare}}\). Black diamonds
are the signals measured on Kingston, with 95\% intervals (hidden by the
symbol in {\srsans\bfseries {a}}).}
\end{figure}

\enlargethispage{\baselineskip}
\begin{tkfsubsec}

Hardware run

\end{tkfsubsec}

Both jobs ran on IBM Kingston on 29 October 2026 (UTC), with four
alternating oracle and control pairs of 5,000 shots each (20,000 shots
per arm), on the same physical qubits as the noisy simulation (Table 2).
Both comparisons met the matching rule, with 228 routed two-qubit gates
in each arm at \(H=2\) and 1,168 and 1,166 at \(H=3\). IBM's job records
retain the submitted circuits and the raw shots of every PUB; the counts
analysed here are identical to those records, and in each job the four
oracle circuits are identical, as are the four controls. At \(H=2\) the
oracle circuit drew a rare future with probability 0.6836 and the
control with 0.4351, a signal of +0.2485 (95\% interval +0.2391 to
+0.2579; \(z=50.1\)), against +0.408 predicted and +0.65 ideal. At
\(H=3\) the probabilities were 0.7353 and 0.7583, a signal of −0.0230
(95\% interval −0.0315 to −0.0144; \(z=-5.3\)), against +0.060 predicted
and +0.52 ideal. By the stated test, the two-step difference reaches the
threshold of \(z\ge5\) set before the runs and the three-step difference
does not reach it.

The following analyses were not specified before the run. The four pairs
gave signals from +0.2418 to +0.2564 at \(H=2\) and from −0.0268 to
−0.0186 at \(H=3\), consistent within sampling error
(\(\chi^2_{3}=1.8\), \(p=0.61\), and \(\chi^2_{3}=0.5\), \(p=0.92\)).
Across the 25 two-step futures, the change in each future's frequency
from control to oracle tracked the change predicted by ideal
amplification (Pearson \(r=0.98\); 0.81 over the 21 rare futures), and
did so within each pair separately (\(r\) from 0.98 to 0.99; 0.74 to
0.84 over rare futures), so the two-step difference has the pattern of
amplification, not only a higher rare-hit rate; across the 125
three-step futures the same correlation was \(r=0.04\).

At two steps the measured signal was 61\% of its prediction, almost
entirely because the control drew rare futures more often than predicted
(0.435 against 0.300; oracle 0.684 against 0.708), consistent with
stronger mixing than the noise model predicted, since the control's
output, concentrated on the most likely future, is the more sensitive to
mixing. Among the shots that returned a rare future, the spread over the
21 rare futures lay 0.141 from the target \(P(x\mid\mathcal R_\Delta)\)
in total variation for the oracle arm and 0.175 for the control, against
0.013 and 0.017 expected from sampling alone and 0.439 for a uniform
spread. Noise therefore distorted the relative likelihoods in both arms
by far more than sampling error, but the amplified arm was no further
from the target than its control.

At three steps the signal had the opposite sign to its prediction in all
four pairs. Both arms drew rare futures far more often than
\(p_{\mathrm{rare}}=0.086\), towards the 0.852 of a uniform output over
all 128 indices (relaxation towards the all-zero index, a non-rare
future, pulls the other way), and their spreads over rare futures (both
0.368 from the target, against 0.029 from sampling alone and 0.457 for a
uniform spread) had lost most of the target's shape. The three-step
outputs were therefore largely noise-dominated. Were both arms
depolarised equally, the oracle arm would still lie above the control,
so the consistent deficit of 0.0230 shows that the two compiled circuits
responded differently to noise: the control departed further from its
prediction (0.758 against 0.607) than the oracle (0.735 against 0.667).
The submitted circuits, retrieved from IBM's job records and simulated
without noise (400,000 shots each), reproduce the ideal values within
sampling error (0.607 and 0.086 at \(H=3\), against 0.607 and 0.086;
0.772 and 0.122 at \(H=2\), against 0.773 and 0.123), confirming that
the deficit does not arise from the submitted circuit logic; its
physical mechanism was not identified. The noise models were optimistic
at both horizons, more so for the deeper circuit. They used calibration
data retrieved at least 6.9 hours before the jobs ran, Kingston was
recalibrated 19 minutes before they ran, and they omit crosstalk,
coherent errors, leakage and drift.

\enlargethispage{\baselineskip}
\begin{tkftabletitle}

{\srsans\bfseries {Table 2.}} Hardware run on IBM Kingston

\end{tkftabletitle}

\begin{longtable}[]{@{}
  >{\raggedright\arraybackslash}p{(\columnwidth - 20\tabcolsep) * \real{0.0694}}
  >{\raggedleft\arraybackslash}p{(\columnwidth - 20\tabcolsep) * \real{0.0556}}
  >{\raggedleft\arraybackslash}p{(\columnwidth - 20\tabcolsep) * \real{0.0972}}
  >{\raggedleft\arraybackslash}p{(\columnwidth - 20\tabcolsep) * \real{0.0972}}
  >{\raggedleft\arraybackslash}p{(\columnwidth - 20\tabcolsep) * \real{0.0972}}
  >{\raggedleft\arraybackslash}p{(\columnwidth - 20\tabcolsep) * \real{0.0972}}
  >{\raggedleft\arraybackslash}p{(\columnwidth - 20\tabcolsep) * \real{0.0972}}
  >{\raggedleft\arraybackslash}p{(\columnwidth - 20\tabcolsep) * \real{0.0972}}
  >{\raggedleft\arraybackslash}p{(\columnwidth - 20\tabcolsep) * \real{0.0972}}
  >{\raggedleft\arraybackslash}p{(\columnwidth - 20\tabcolsep) * \real{0.0972}}
  >{\raggedleft\arraybackslash}p{(\columnwidth - 20\tabcolsep) * \real{0.0972}}@{}}
\toprule\noalign{}
\begin{minipage}[b]{\linewidth}\raggedright
\(H\)
\end{minipage} & \begin{minipage}[b]{\linewidth}\raggedleft
\(r\)
\end{minipage} & \begin{minipage}[b]{\linewidth}\raggedleft
Shots per arm
\end{minipage} & \begin{minipage}[b]{\linewidth}\raggedleft
Routed 2q, oracle / control
\end{minipage} & \begin{minipage}[b]{\linewidth}\raggedleft
\(P(\mathrm{rare})\), oracle
\end{minipage} & \begin{minipage}[b]{\linewidth}\raggedleft
\(P(\mathrm{rare})\), control
\end{minipage} & \begin{minipage}[b]{\linewidth}\raggedleft
Signal
\end{minipage} & \begin{minipage}[b]{\linewidth}\raggedleft
95\% interval
\end{minipage} & \begin{minipage}[b]{\linewidth}\raggedleft
\(z\)
\end{minipage} & \begin{minipage}[b]{\linewidth}\raggedleft
Predicted signal
\end{minipage} & \begin{minipage}[b]{\linewidth}\raggedleft
Ideal signal
\end{minipage} \\
\midrule\noalign{}
\endhead
\bottomrule\noalign{}
\endlastfoot
2 & 1 & 20,000 & 228 / 228 & 0.6836 & 0.4351 & +0.2485 & {[}+0.2391,
+0.2579{]} & 50.1 & +0.408 & +0.650 \\
3 & 1 & 20,000 & 1,168 / 1,166 & 0.7353 & 0.7583 & −0.0230 & {[}−0.0315,
−0.0144{]} & −5.3 & +0.060 & +0.521 \\
\end{longtable}

\begin{tkftablenote}

Four oracle and control pairs of 5,000 shots each per job, submitted
alternately. \(z\) is the pooled two-proportion statistic of the stated
test; the interval uses the unpooled standard error. Predicted signal is
the mean of the Kingston noisy simulation (Supplementary Table S4). Jobs
dbhgjlht9uhs73ca7e9g and dbhgk2hq260s73bv9100. Calibration in force at
run time on the qubits used: \(H=2\): qubits 3, 24, 16, 22, 23, median
CZ error \(1.2\times10^{-3}\), readout error \(1.2\times10^{-2}\),
\(T_1\) 227 µs, \(T_2\) 118 µs. \(H=3\): qubits 75, 71, 93, 79, 74, 72,
73, median CZ error \(1.2\times10^{-3}\), readout error
\(1.3\times10^{-2}\), \(T_1\) 248 µs, \(T_2\) 154 µs.

\end{tkftablenote}

\vspace{-12pt}

\begin{tkfsec}

Discussion

\end{tkfsec}

The pipeline shows that, when the process is a known finite-state Markov
chain, the rare set defined as in Guo et~al.~(2026) can be marked by a
circuit that is exactly verifiable at this scale. This differs from
their setting, in which the marking is built from the preparation
circuit alone. The circuit combines standard techniques: surprisal
accumulated in the Fourier basis (Draper, 2000; Ruiz-Perez \&
Garcia-Escartin, 2017), so each transition costs one diagonal gate per
accumulator qubit; an offset preload, so the comparison reduces to a
single-qubit phase; and reliance on the prepared state's support, so
unused encodings need no flag. The validation separates circuit errors,
of which there were none, from precision errors, which are a property of
the fixed-point definition and are reported as probability mass.

Exact marking has a cost that does not follow the horizon smoothly.
Guaranteed classification needs \(b\) of about \(\log_2[(H/2+1)/g]\),
where \(g\) is the gap between the threshold and the nearest future, and
\(g\) varies irregularly with the horizon (Supplementary Table S1): the
bound is 15 at \(H=5\) and 13 at \(H=6\). Below it, zero mismatches
depend on how the rounding errors fall and must be checked for each
\(b\) (Supplementary Table S2).

The results have four limits. First, the circuit uses the transition
matrix, so the same information gives exact classical sampling of the
rare-conditioned distribution; the access model differs from the
sample-access comparison of Guo et~al.~(2026), and no advantage is
claimed. Second, the six-step results are ideal simulations; the
scalable circuit, at about 66,802 logical two-qubit gates for four
iterations at \(b=7\), is far beyond present hardware. Third, the
hardware test uses non-scalable references with the rare set supplied as
a list, so it tests amplification and the noise model, not the scalable
oracle; a passing test shows that oracle and control behave differently
in the predicted direction, not that every property of amplification
survives noise, and the three-step run shows that the two circuits can
also differ under noise for reasons unrelated to amplification; the
per-future pattern supports amplification at two steps, and a noiseless
replay of the submitted three-step circuits confirms their intended
ideal behaviour. Fourth, the noise models omit crosstalk, coherent
errors, leakage and drift; on Kingston the two-step signal was 61\% of
its prediction, and the three-step prediction of a detectable signal was
not borne out.

The mechanism is aimed at models that can be run but whose probabilities
are hard to calculate, such as large climate, epidemic and financial
simulators whose observed paths aggregate many hidden ones. There the
rule-based oracle demonstrated here cannot be built; it would be
replaced by a rarity test built from the generator alone, as in Guo
et~al.~(2026). That test needs the simulator itself, and its inverse, to
run coherently as a quantum circuit, which is the dominant cost and
remains open for simulators of this size.

\begin{tkfsec}

Conclusion

\end{tkfsec}

A gate-level Qiskit circuit marks and amplifies the rare futures of a
five-state Markov process over six steps, with every component verified
against exact enumeration. At 7 fractional bits it marks the real-valued
rare set exactly, and its 30-qubit simulation raises the probability of
a rare future from 0.025 to 0.980 in four iterations. On IBM Kingston, a
two-step reference circuit of 228 two-qubit gates, which marks rare
futures from a precomputed list, drew rare futures more often than its
matched control by +0.2485 (\(z=50.1\)), passing the test set before the
run, while at three steps, with 1,168 two-qubit gates, the signal
reversed sign. A hardware signature consistent with amplitude
amplification was therefore detected on Kingston at 228 two-qubit gates
but not at 1,168, and the complete six-step circuit, at tens of
thousands, awaits substantially higher-fidelity quantum hardware.

\enlargethispage{\baselineskip}

\begin{tkfsec}

Acknowledgements

\end{tkfsec}

We acknowledge the use of IBM Quantum services for this work. The
hardware experiments ran on ibm\_kingston, a QPU built on an IBM Heron
r2 processor. We thank the IBM Quantum team for this access and for
Qiskit, and look forward to the higher-fidelity processors that will
bring the six-step circuit within reach. The views expressed are those
of the author, and do not reflect the official policy or position of IBM
or the IBM Quantum team.

\clearpage

\begin{tkfsec}

References

\end{tkfsec}

\begin{tkfref}

Asmussen, S., \& Glynn, P. W. (2007). \emph{Stochastic simulation:
Algorithms and analysis}. Springer.
\url{https://doi.org/10.1007/978-0-387-69033-9}

\end{tkfref}

\begin{tkfref}

Brassard, G., Høyer, P., Mosca, M., \& Tapp, A. (2002). Quantum
amplitude amplification and estimation. \emph{Contemporary Mathematics},
\emph{305}, 53--74. \url{https://doi.org/10.1090/conm/305/05215}

\end{tkfref}

\begin{tkfref}

Bucklew, J. A. (2004). \emph{Introduction to rare event simulation}.
Springer. \url{https://doi.org/10.1007/978-1-4757-4078-3}

\end{tkfref}

\begin{tkfref}

Catt, P. M. (2026). \emph{Forecastability profiles: Horizon-specific
limits to prediction before model selection} {[}Preprint{]}. Research
Square. \url{https://doi.org/10.21203/rs.3.rs-9882043/v1}

\end{tkfref}

\begin{tkfref}

Cérou, F., \& Guyader, A. (2007). Adaptive multilevel splitting for rare
event analysis. \emph{Stochastic Analysis and Applications},
\emph{25}(2), 417--443. \url{https://doi.org/10.1080/07362990601139628}

\end{tkfref}

\begin{tkfref}

Cover, T. M., \& Thomas, J. A. (2006). \emph{Elements of information
theory} (2nd ed.). Wiley. \url{https://doi.org/10.1002/047174882X}

\end{tkfref}

\begin{tkfref}

Del Moral, P., \& Garnier, J. (2005). Genealogical particle analysis of
rare events. \emph{The Annals of Applied Probability}, \emph{15}(4),
2496--2534. \url{https://doi.org/10.1214/105051605000000566}

\end{tkfref}

\begin{tkfref}

Dobson, I., Carreras, B. A., Lynch, V. E., \& Newman, D. E. (2007).
Complex systems analysis of series of blackouts: Cascading failure,
critical points, and self-organization. \emph{Chaos}, \emph{17}(2),
Article 026103. \url{https://doi.org/10.1063/1.2737822}

\end{tkfref}

\begin{tkfref}

Draper, T. G. (2000). \emph{Addition on a quantum computer}
(arXiv:quant-ph/0008033). arXiv.
\url{https://doi.org/10.48550/arXiv.quant-ph/0008033}

\end{tkfref}

\begin{tkfref}

Dupuis, P., \& Wang, H. (2005). Dynamic importance sampling for
uniformly recurrent Markov chains. \emph{The Annals of Applied
Probability}, \emph{15}(1A), 1--38.
\url{https://doi.org/10.1214/105051604000001016}

\end{tkfref}

\begin{tkfref}

Egger, D. J., García Gutiérrez, R., Cahué Mestre, J., \& Woerner, S.
(2021). Credit risk analysis using quantum computers. \emph{IEEE
Transactions on Computers}, \emph{70}(12), 2136--2145.
\url{https://doi.org/10.1109/TC.2020.3038063}

\end{tkfref}

\begin{tkfref}

Figgatt, C., Maslov, D., Landsman, K. A., Linke, N. M., Debnath, S., \&
Monroe, C. (2017). Complete 3-qubit Grover search on a programmable
quantum computer. \emph{Nature Communications}, \emph{8}, Article 1918.
\url{https://doi.org/10.1038/s41467-017-01904-7}

\end{tkfref}

\begin{tkfref}

Grover, L. K. (1996). A fast quantum mechanical algorithm for database
search. In \emph{Proceedings of the 28th Annual ACM Symposium on Theory
of Computing} (pp.~212--219). Association for Computing Machinery.
\url{https://doi.org/10.1145/237814.237866}

\end{tkfref}

\begin{tkfref}

Grover, L., \& Rudolph, T. (2002). \emph{Creating superpositions that
correspond to efficiently integrable probability distributions}
(arXiv:quant-ph/0208112). arXiv.
\url{https://doi.org/10.48550/arXiv.quant-ph/0208112}

\end{tkfref}

\begin{tkfref}

Guo, N., Huang, P.-W., Wang, Q., Thompson, J., Rebentrost, P., Gu, M.,
\& Yang, C. (2026). \emph{Quantum enhanced rare event discovery and
sampling} (arXiv:2606.06316). arXiv.
\url{https://doi.org/10.48550/arXiv.2606.06316}

\end{tkfref}

\begin{tkfref}

Haldane, A. G., \& May, R. M. (2011). Systemic risk in banking
ecosystems. \emph{Nature}, \emph{469}(7330), 351--355.
\url{https://doi.org/10.1038/nature09659}

\end{tkfref}

\begin{tkfref}

Javadi-Abhari, A., Treinish, M., Krsulich, K., Wood, C. J., Lishman, J.,
Gacon, J., Martiel, S., Nation, P. D., Bishop, L. S., Cross, A. W.,
Johnson, B. R., \& Gambetta, J. M. (2024). \emph{Quantum computing with
Qiskit} (arXiv:2405.08810). arXiv.
\url{https://doi.org/10.48550/arXiv.2405.08810}

\end{tkfref}

\begin{tkfref}

Montanaro, A. (2015). Quantum speedup of Monte Carlo methods.
\emph{Proceedings of the Royal Society A}, \emph{471}(2181), Article
20150301. \url{https://doi.org/10.1098/rspa.2015.0301}

\end{tkfref}

\begin{tkfref}

Möttönen, M., Vartiainen, J. J., Bergholm, V., \& Salomaa, M. M. (2004).
Quantum circuits for general multiqubit gates. \emph{Physical Review
Letters}, \emph{93}(13), Article 130502.
\url{https://doi.org/10.1103/PhysRevLett.93.130502}

\end{tkfref}

\begin{tkfref}

Möttönen, M., Vartiainen, J. J., Bergholm, V., \& Salomaa, M. M. (2005).
Transformation of quantum states using uniformly controlled rotations.
\emph{Quantum Information \& Computation}, \emph{5}(6), 467--473.
\url{https://doi.org/10.26421/QIC5.6-5}

\end{tkfref}

\begin{tkfref}

Ruiz-Perez, L., \& Garcia-Escartin, J. C. (2017). Quantum arithmetic
with the quantum Fourier transform. \emph{Quantum Information
Processing}, \emph{16}(6), Article 152.
\url{https://doi.org/10.1007/s11128-017-1603-1}

\end{tkfref}

\begin{tkfref}

Shende, V. V., Bullock, S. S., \& Markov, I. L. (2006). Synthesis of
quantum-logic circuits. \emph{IEEE Transactions on Computer-Aided Design
of Integrated Circuits and Systems}, \emph{25}(6), 1000--1010.
\url{https://doi.org/10.1109/TCAD.2005.855930}

\end{tkfref}

\begin{tkfref}

Stamatopoulos, N., Egger, D. J., Sun, Y., Zoufal, C., Iten, R., Shen,
N., \& Woerner, S. (2020). Option pricing using quantum computers.
\emph{Quantum}, \emph{4}, Article 291.
\url{https://doi.org/10.22331/q-2020-07-06-291}

\end{tkfref}

\begin{tkfref}

Vidal, G. (2003). Efficient classical simulation of slightly entangled
quantum computations. \emph{Physical Review Letters}, \emph{91}(14),
Article 147902. \url{https://doi.org/10.1103/PhysRevLett.91.147902}

\end{tkfref}

\begin{tkfref}

Woerner, S., \& Egger, D. J. (2019). Quantum risk analysis. \emph{npj
Quantum Information}, \emph{5}, Article 15.
\url{https://doi.org/10.1038/s41534-019-0130-6}

\end{tkfref}

\begin{tkfref}

Zscheischler, J., Westra, S., van den Hurk, B. J. J. M., Seneviratne, S.
I., Ward, P. J., Pitman, A., AghaKouchak, A., Bresch, D. N., Leonard,
M., Wahl, T., \& Zhang, X. (2018). Future climate risk from compound
events. \emph{Nature Climate Change}, \emph{8}(6), 469--477.
\url{https://doi.org/10.1038/s41558-018-0156-3}

\end{tkfref}

\clearpage

\begin{tkfappendix}

Supplementary Information

\end{tkfappendix}

\begin{tkftabletitle}

{\srsans\bfseries {Supplementary Table S1.}} Rare sets for horizons 2 to
6 (\(\delta=1.5\), start state 1)

\end{tkftabletitle}

\begin{longtable}[]{@{}
  >{\raggedright\arraybackslash}p{(\columnwidth - 14\tabcolsep) * \real{0.0926}}
  >{\raggedleft\arraybackslash}p{(\columnwidth - 14\tabcolsep) * \real{0.1296}}
  >{\raggedleft\arraybackslash}p{(\columnwidth - 14\tabcolsep) * \real{0.1296}}
  >{\raggedleft\arraybackslash}p{(\columnwidth - 14\tabcolsep) * \real{0.1296}}
  >{\raggedleft\arraybackslash}p{(\columnwidth - 14\tabcolsep) * \real{0.1296}}
  >{\raggedleft\arraybackslash}p{(\columnwidth - 14\tabcolsep) * \real{0.1296}}
  >{\raggedleft\arraybackslash}p{(\columnwidth - 14\tabcolsep) * \real{0.1296}}
  >{\raggedleft\arraybackslash}p{(\columnwidth - 14\tabcolsep) * \real{0.1296}}@{}}
\toprule\noalign{}
\begin{minipage}[b]{\linewidth}\raggedright
\(H\)
\end{minipage} & \begin{minipage}[b]{\linewidth}\raggedleft
Futures
\end{minipage} & \begin{minipage}[b]{\linewidth}\raggedleft
Rare
\end{minipage} & \begin{minipage}[b]{\linewidth}\raggedleft
\(p_{\mathrm{rare}}\)
\end{minipage} & \begin{minipage}[b]{\linewidth}\raggedleft
\(r^*\)
\end{minipage} & \begin{minipage}[b]{\linewidth}\raggedleft
Ideal success
\end{minipage} & \begin{minipage}[b]{\linewidth}\raggedleft
Nearest future to threshold (bits)
\end{minipage} & \begin{minipage}[b]{\linewidth}\raggedleft
\(b\) for zero mismatches
\end{minipage} \\
\midrule\noalign{}
\endhead
\bottomrule\noalign{}
\endlastfoot
2 & 25 & 21 & 0.1227 & 2 & 0.953 & \(6.4\times10^{-2}\) & 1 \\
3 & 125 & 109 & 0.0860 & 2 & 0.993 & \(2.0\times10^{-2}\) & 2 \\
4 & 625 & 527 & 0.0453 & 3 & 0.995 & \(9.6\times10^{-3}\) & 3 \\
5 & 3,125 & 2,731 & 0.0375 & 4 & 0.967 & \(1.4\times10^{-4}\) & 13 \\
6 & 15,625 & 13,888 & 0.0250 & 4 & 0.980 & \(5.7\times10^{-4}\) & 7 \\
\end{longtable}

\begin{tkftablenote}

\(r^*=\lfloor\pi/(4\theta)\rfloor\) and ideal success is
\(\sin^2[(2r^*+1)\theta]\). The last column is the lowest \(b\) found by
checking all futures. Rare futures are the majority by count at every
horizon but carry little of the probability.

\end{tkftablenote}

\begin{tkftabletitle}

{\srsans\bfseries {Supplementary Table S2.}} Fixed-point precision at
\(H=6\)

\end{tkftabletitle}

\begin{longtable}[]{@{}
  >{\raggedright\arraybackslash}p{(\columnwidth - 12\tabcolsep) * \real{0.1064}}
  >{\raggedleft\arraybackslash}p{(\columnwidth - 12\tabcolsep) * \real{0.1489}}
  >{\raggedleft\arraybackslash}p{(\columnwidth - 12\tabcolsep) * \real{0.1489}}
  >{\raggedleft\arraybackslash}p{(\columnwidth - 12\tabcolsep) * \real{0.1489}}
  >{\raggedleft\arraybackslash}p{(\columnwidth - 12\tabcolsep) * \real{0.1489}}
  >{\raggedleft\arraybackslash}p{(\columnwidth - 12\tabcolsep) * \real{0.1489}}
  >{\raggedleft\arraybackslash}p{(\columnwidth - 12\tabcolsep) * \real{0.1489}}@{}}
\toprule\noalign{}
\begin{minipage}[b]{\linewidth}\raggedright
\(b\)
\end{minipage} & \begin{minipage}[b]{\linewidth}\raggedleft
Threshold \(T_b\)
\end{minipage} & \begin{minipage}[b]{\linewidth}\raggedleft
Accumulator qubits \(m\)
\end{minipage} & \begin{minipage}[b]{\linewidth}\raggedleft
Offset \(c\)
\end{minipage} & \begin{minipage}[b]{\linewidth}\raggedleft
Total qubits
\end{minipage} & \begin{minipage}[b]{\linewidth}\raggedleft
Mismatched futures
\end{minipage} & \begin{minipage}[b]{\linewidth}\raggedleft
Misclassified mass
\end{minipage} \\
\midrule\noalign{}
\endhead
\bottomrule\noalign{}
\endlastfoot
1 & 32 & 6 & 0 & 24 & 144 & \(2.3\times10^{-3}\) \\
2 & 64 & 7 & 0 & 25 & 60 & \(9.4\times10^{-4}\) \\
3 & 128 & 8 & 0 & 26 & 82 & \(1.3\times10^{-3}\) \\
4 & 255 & 9 & 1 & 27 & 9 & \(1.4\times10^{-4}\) \\
5 & 510 & 10 & 2 & 28 & 12 & \(2.0\times10^{-4}\) \\
6 & 1,019 & 11 & 5 & 29 & 37 & \(6.0\times10^{-4}\) \\
7 & 2,038 & 12 & 10 & 30 & 0 & 0 \\
8 & 4,076 & 13 & 20 & 31 & 0 & 0 \\
9 & 8,152 & 14 & 40 & 32 & 2 & \(3.2\times10^{-5}\) \\
10 & 16,304 & 15 & 80 & 33 & 0 & 0 \\
\end{longtable}

\begin{tkftablenote}

Mismatches compare the fixed-point test with the real-valued rare-set
definition over all 15,625 futures.

\end{tkftablenote}

\begin{tkftabletitle}

{\srsans\bfseries {Supplementary Table S3.}} Logical cost of the
scalable circuit (all-to-all connectivity)

\end{tkftabletitle}

\begin{longtable}[]{@{}
  >{\raggedright\arraybackslash}p{(\columnwidth - 14\tabcolsep) * \real{0.0893}}
  >{\raggedright\arraybackslash}p{(\columnwidth - 14\tabcolsep) * \real{0.1607}}
  >{\raggedleft\arraybackslash}p{(\columnwidth - 14\tabcolsep) * \real{0.1250}}
  >{\raggedleft\arraybackslash}p{(\columnwidth - 14\tabcolsep) * \real{0.1250}}
  >{\raggedleft\arraybackslash}p{(\columnwidth - 14\tabcolsep) * \real{0.1250}}
  >{\raggedleft\arraybackslash}p{(\columnwidth - 14\tabcolsep) * \real{0.1250}}
  >{\raggedleft\arraybackslash}p{(\columnwidth - 14\tabcolsep) * \real{0.1250}}
  >{\raggedleft\arraybackslash}p{(\columnwidth - 14\tabcolsep) * \real{0.1250}}@{}}
\toprule\noalign{}
\begin{minipage}[b]{\linewidth}\raggedright
\(H\)
\end{minipage} & \begin{minipage}[b]{\linewidth}\raggedright
Precision
\end{minipage} & \begin{minipage}[b]{\linewidth}\raggedleft
\(b\)
\end{minipage} & \begin{minipage}[b]{\linewidth}\raggedleft
Qubits
\end{minipage} & \begin{minipage}[b]{\linewidth}\raggedleft
Oracle 2q
\end{minipage} & \begin{minipage}[b]{\linewidth}\raggedleft
Iteration 2q
\end{minipage} & \begin{minipage}[b]{\linewidth}\raggedleft
Iterations
\end{minipage} & \begin{minipage}[b]{\linewidth}\raggedleft
Total 2q
\end{minipage} \\
\midrule\noalign{}
\endhead
\bottomrule\noalign{}
\endlastfoot
2 & fixed & 4 & 14 & 2,441 & 2,593 & 2 & 5,248 \\
2 & exact & 1 & 11 & 1,453 & 1,605 & 2 & 3,272 \\
3 & fixed & 4 & 17 & 4,437 & 4,725 & 2 & 9,568 \\
3 & exact & 2 & 15 & 3,309 & 3,597 & 2 & 7,312 \\
4 & fixed & 4 & 21 & 7,281 & 7,705 & 3 & 23,289 \\
4 & exact & 3 & 20 & 6,457 & 6,881 & 3 & 20,817 \\
5 & fixed & 4 & 24 & 9,529 & 10,089 & 4 & 40,586 \\
5 & exact & 13 & 33 & 19,725 & 20,283 & 4 & 81,362 \\
6 & fixed & 4 & 27 & 11,777 & 12,473 & 4 & 50,178 \\
6 & exact & 7 & 30 & 15,933 & 16,629 & 4 & 66,802 \\
\end{longtable}

\begin{tkftablenote}

Basis \(\{R_z,\sqrt X,X,\mathrm{CNOT}\}\), optimisation level 3. Fixed
precision is \(b=4\); exact precision is the lowest \(b\) with zero
mismatches. Total counts one preparation plus the stated number of
iterations.

\end{tkftablenote}

\begin{tkftabletitle}

{\srsans\bfseries {Supplementary Table S4.}} Predicted
oracle-minus-control signal (noise models from IBM calibration snapshots
retrieved on 29 October 2026)

\end{tkftabletitle}

\begin{longtable}[]{@{}
  >{\raggedright\arraybackslash}p{(\columnwidth - 16\tabcolsep) * \real{0.0847}}
  >{\raggedleft\arraybackslash}p{(\columnwidth - 16\tabcolsep) * \real{0.0678}}
  >{\raggedright\arraybackslash}p{(\columnwidth - 16\tabcolsep) * \real{0.1356}}
  >{\raggedleft\arraybackslash}p{(\columnwidth - 16\tabcolsep) * \real{0.1186}}
  >{\raggedleft\arraybackslash}p{(\columnwidth - 16\tabcolsep) * \real{0.1186}}
  >{\raggedleft\arraybackslash}p{(\columnwidth - 16\tabcolsep) * \real{0.1186}}
  >{\raggedleft\arraybackslash}p{(\columnwidth - 16\tabcolsep) * \real{0.1186}}
  >{\raggedleft\arraybackslash}p{(\columnwidth - 16\tabcolsep) * \real{0.1186}}
  >{\raggedleft\arraybackslash}p{(\columnwidth - 16\tabcolsep) * \real{0.1186}}@{}}
\toprule\noalign{}
\begin{minipage}[b]{\linewidth}\raggedright
\(H\)
\end{minipage} & \begin{minipage}[b]{\linewidth}\raggedleft
\(r\)
\end{minipage} & \begin{minipage}[b]{\linewidth}\raggedright
Backend
\end{minipage} & \begin{minipage}[b]{\linewidth}\raggedleft
Routed 2q, oracle / control
\end{minipage} & \begin{minipage}[b]{\linewidth}\raggedleft
Ideal \(P(\mathrm{rare})\)
\end{minipage} & \begin{minipage}[b]{\linewidth}\raggedleft
Noisy, oracle
\end{minipage} & \begin{minipage}[b]{\linewidth}\raggedleft
Noisy, control
\end{minipage} & \begin{minipage}[b]{\linewidth}\raggedleft
Signal
\end{minipage} & \begin{minipage}[b]{\linewidth}\raggedleft
\(z\)
\end{minipage} \\
\midrule\noalign{}
\endhead
\bottomrule\noalign{}
\endlastfoot
2 & 1 & Fez & 228 / 228 & 0.773 & 0.700 & 0.344 & +0.356 & 71.3 \\
2 & 1 & Kingston & 228 / 228 & 0.773 & 0.708 & 0.300 & +0.408 & 81.7 \\
2 & 1 & Marrakesh & 228 / 228 & 0.773 & 0.698 & 0.346 & +0.352 & 70.4 \\
2 & 2 & Fez & 440 / 440 & 0.953 & 0.763 & 0.467 & +0.296 & 60.9 \\
2 & 2 & Kingston & 440 / 440 & 0.953 & 0.787 & 0.431 & +0.356 & 73.0 \\
2 & 2 & Marrakesh & 440 / 440 & 0.953 & 0.758 & 0.468 & +0.290 & 59.5 \\
3 & 1 & Fez † & 1,168 / 1,114 & 0.607 & 0.719 & 0.673 & +0.046 & 10.0 \\
3 & 1 & Kingston & 1,168 / 1,166 & 0.607 & 0.667 & 0.607 & +0.060 &
12.5 \\
3 & 1 & Marrakesh † & 1,168 / 1,114 & 0.607 & 0.717 & 0.678 & +0.039 &
8.6 \\
3 & 2 & Fez † & 2,122 / 2,020 & 0.993 & 0.804 & 0.785 & +0.019 & 4.7 \\
3 & 2 & Kingston † & 2,122 / 2,020 & 0.993 & 0.783 & 0.738 & +0.045 &
10.7 \\
3 & 2 & Marrakesh † & 2,122 / 2,020 & 0.993 & 0.805 & 0.786 & +0.019 &
4.6 \\
\end{longtable}

\begin{tkftablenote}

† Control and oracle differ by more than 2\% in routed two-qubit count.
Means over 3 repeats of 20,000 shots per arm; \(z\) is the pooled
two-proportion statistic of the stated test. The ideal control
probability equals \(p_{\mathrm{rare}}\): 0.1227 at \(H=2\) and 0.0860
at \(H=3\). At \(r=0\) the two arms are identical circuits; the largest
mean difference, \(3.3\times10^{-3}\), is consistent with sampling
error.

\end{tkftablenote}

\end{document}